%% file: main.tex
\documentclass[11pt,titlepage]{article}
\usepackage[utf8]{inputenc}
\usepackage{geometry}
\usepackage{placeins}
\usepackage{float}
\usepackage{graphicx,wrapfig}
\usepackage{amsmath}
\usepackage{caption}
\usepackage{subcaption}
\usepackage[dvipsnames]{xcolor} 

\usepackage{jinstpub} 

\usepackage{siunitx}

\date{} 

\author[1]{N.~Akchurin,}
\author[1]{J.~Damgov,}
\author[2]{S.~Dugad,}
\author[1]{P.~G C,}
\author[3]{S.~Gr\"onroos,}
\author[1]{K.~Lamichhane}
\author[1]{J.~Martinez,}
\author[3]{T.~Quast,}
\author[1]{S.~Undleeb,}
\author[1]{A.~Whitbeck}

\affiliation[1]{Advanced Particle Detector Laboratory, Department of Physics and Astronomy, Texas Tech University,  {Lubbock}, TX, 79409, USA}

\affiliation[2]{Tata Institute of Fundamental Research, Colaba, Mumbai, India}

\affiliation[3]{CERN, Experimental Physics Department, Esplanade des Particules 1, 1211 Meyrin, CH}
\emailAdd{kamal.lamichhane@ttu.edu, thorben.quast@cern.ch}

\newcommand\snowmass{\begin{center}\rule[-0.2in]{\hsize}{0.01in}\\\rule{\hsize}{0.01in}\\
\vskip 0.1in Submitted to the  Proceedings of the US Community Study\\ 
on the Future of Particle Physics (Snowmass 2021)\\ 
\rule{\hsize}{0.01in}\\\rule[+0.2in]{\hsize}{0.01in} \end{center}}

\title{Deep learning applications for quality control in particle detector construction}

\abstract{The growing complexity of particle detectors makes their construction and quality control a new challenge. We present studies that explore the use of deep learning-based computer vision techniques to perform quality checks of detector components and assembly steps, which will automate procedures and minimize the need for human interventions. This study focuses on the construction
steps of a silicon detector, which involve forming a mechanical structure with the sensor and wire
bonding individual cells to electronics for reading out signals. Silicon detectors in high energy physics
experiments today have millions of channels. Manual quality control of these and other high channel-density detectors requires enormous amounts of labor and can be prone to errors. Here, we explore computer vision applications to either augment or fully replace visual inspections done by humans. We investigated
convolutional neural networks for image classification and autoencoders for anomalies detection. Two proof-of-concept studies will be presented.

\snowmass
}

\begin{document}
\maketitle

\input{intro}

\section{Quality Control using Deep Learning}
\label{sec:qc}
In what follows, image classification using CNNs and anomaly detection using autoencoders (AEs) are presented.  In both cases, images were collected using microscopes attached to programmable gantries.  In the case of module bonding area inspection, a program was developed to automatically locate each of the 86 bonding areas of a given module.  For sensor inspection, images are captured automatically in fixed steps such that a mosaic of the entire sensor could be formed from these images. 

\input{ttu}

\input{cern}

\input{summary}

\newpage
\bibliography{bib.bib}

\end{document}

%% file: intro.tex
\section{Introduction}
\label{sec:intro}

The demands of particle physics experiments are necessitating particle detector designs of increasing complexity.  Such
complexity is often a result of demand on the detector's size, channel density, active materials, operating conditions (such as temperature), or tight mechanical tolerances.  For example, silicon detectors are widely used in high energy physics collider experiments and space-borne experiments.  In collider scenarios,  silicon detectors need to operate at cold temperatures to minimize leakage currents, especially after the large doses of radiation they are exposed to over the lifetime of an experiment.  Low operating temperatures can necessitate strict mechanical tolerances on the system.  High channel densities are necessary for silicon-based tracking detectors to produce high precisions measurements.  In the case of strip sensors, this means millions of wire bonds might be deployed to analyze signals.  In calorimetry, high channel-density silicon detectors are being designed to help mitigate the effects of radiation damage and high pileup conditions, a new challenge necessitated by efforts to explore the rarest processes at hadron colliders.  

In both space-borne and collider experiments, where detectors are not readily accessible or running time is a premium, reliable detector systems are crucial to the success of scientific programs.   Detailed quality control (QC) testing is key to producing high-performance, reliable detectors.  Limited timelines for constructing new experiments intensify assembly work and QC testing necessary to produce reliable detectors.  Visual inspections of large detectors or small features of detectors can be an essential but time-consuming part of the QC process.  Hence, new tools for automating QC are necessary to minimize time and errors as our detectors grow in complexity.   In the inspection of components, computer vision techniques based on deep learning algorithms~\cite{lecun-nature-15} offer many advantages for quick and precise QC.

\begin{figure}[!htb]
     \centering
   
    \begin{subfigure}[b]{0.47\textwidth}
         \centering
         \includegraphics[width=\textwidth]{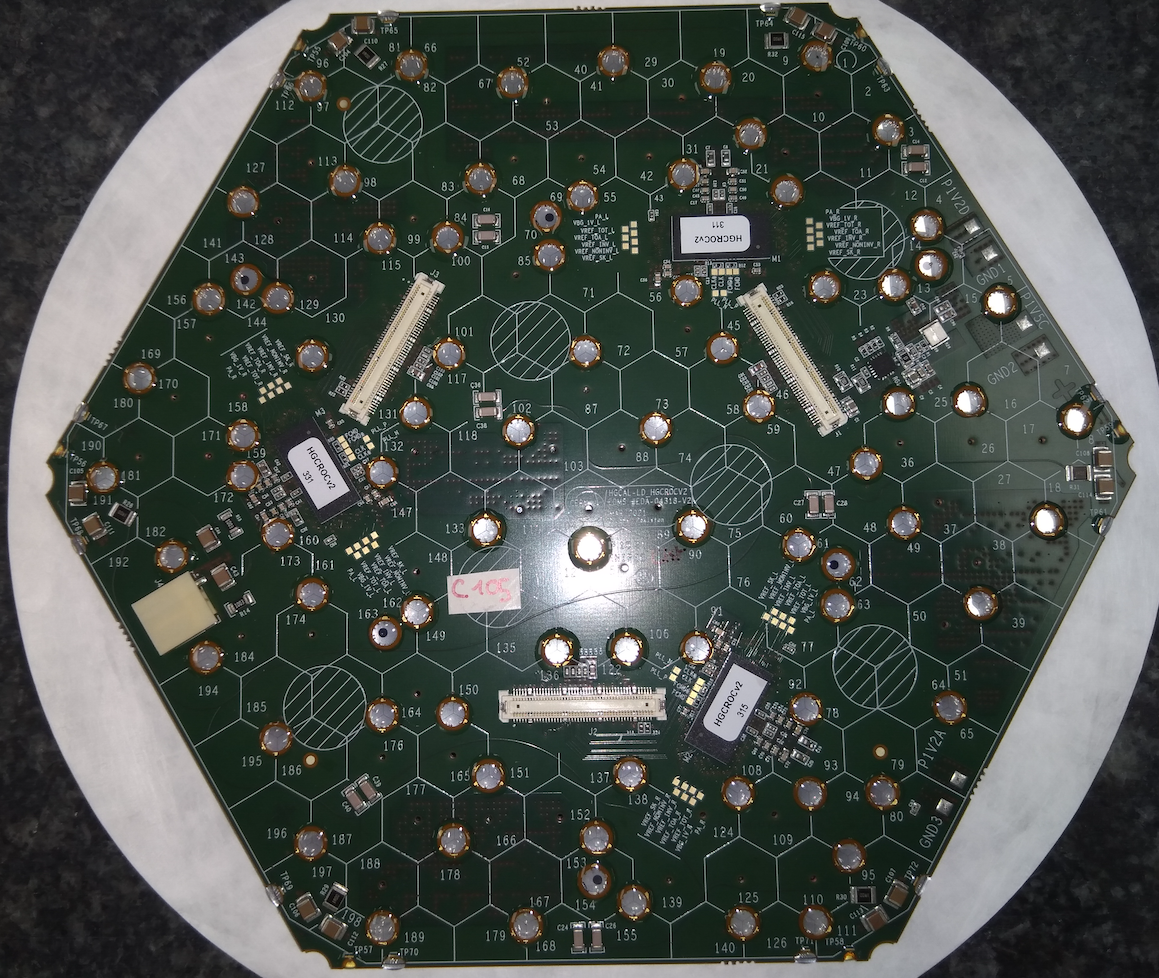}
         \caption{}
         \label{fig:module}
     \end{subfigure}
     \hfill
     \begin{subfigure}[b]{0.4\textwidth}
         \centering
         \includegraphics[width=\textwidth]{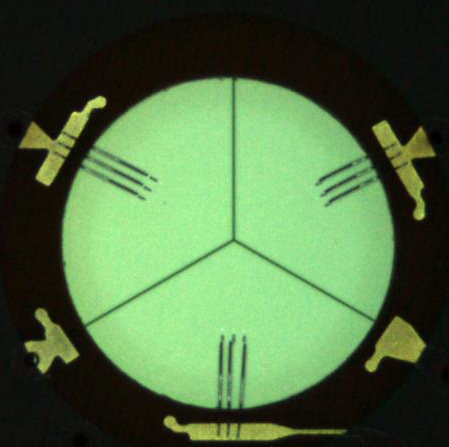}
         \caption{}
         \label{fig:wirebond-img}
     \end{subfigure}
        \caption{ a)  An HGCAL complete module with a circuit board  glued on top of a silicon sensor. Each white hexagon represent a distinct detector channel.  The holes seen at the vertices of the channel boundaries are designed to connect the circuit board and the sensor with wire bonds. b) A zoomed image of a wire bond hole. The black lines emanating from the center of the circle are the edges of a metal layer that mark channel boundaries.  Lines extending from the gold bond pad are individual wires that were bonded.}
        \label{fig:construction}
\end{figure}

\begin{figure}[!htb]
	\centering
	\begin{subfigure}{0.325\textwidth}
		\includegraphics[width=1.\textwidth]{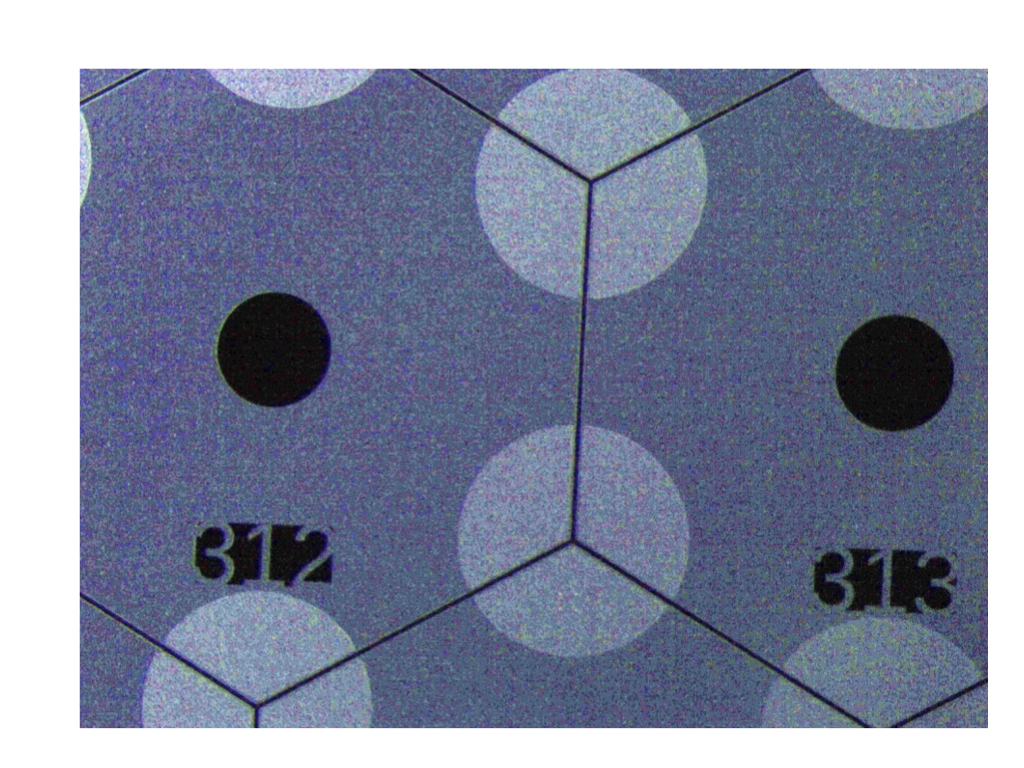}
		\subcaption{}		
	\end{subfigure}
	\hfill
	\begin{subfigure}{0.325\textwidth}
		\includegraphics[width=1.\textwidth]{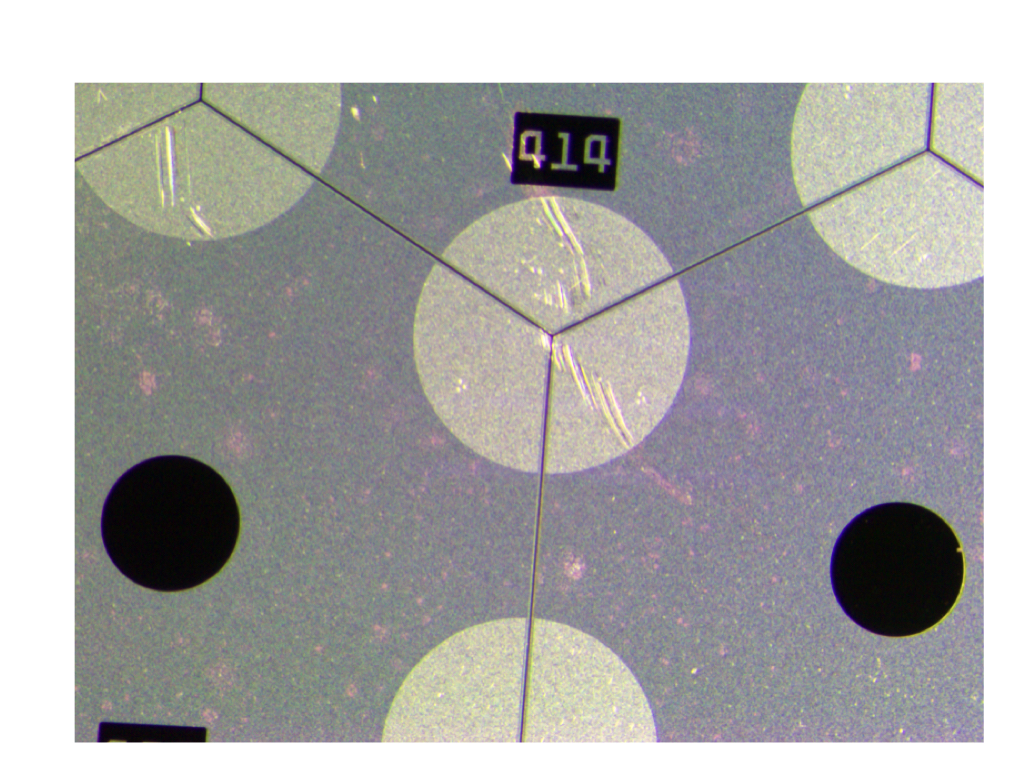}
		\subcaption{}		
	\end{subfigure}
	\hfill
	\begin{subfigure}{0.325\textwidth}
		\includegraphics[width=1.\textwidth]{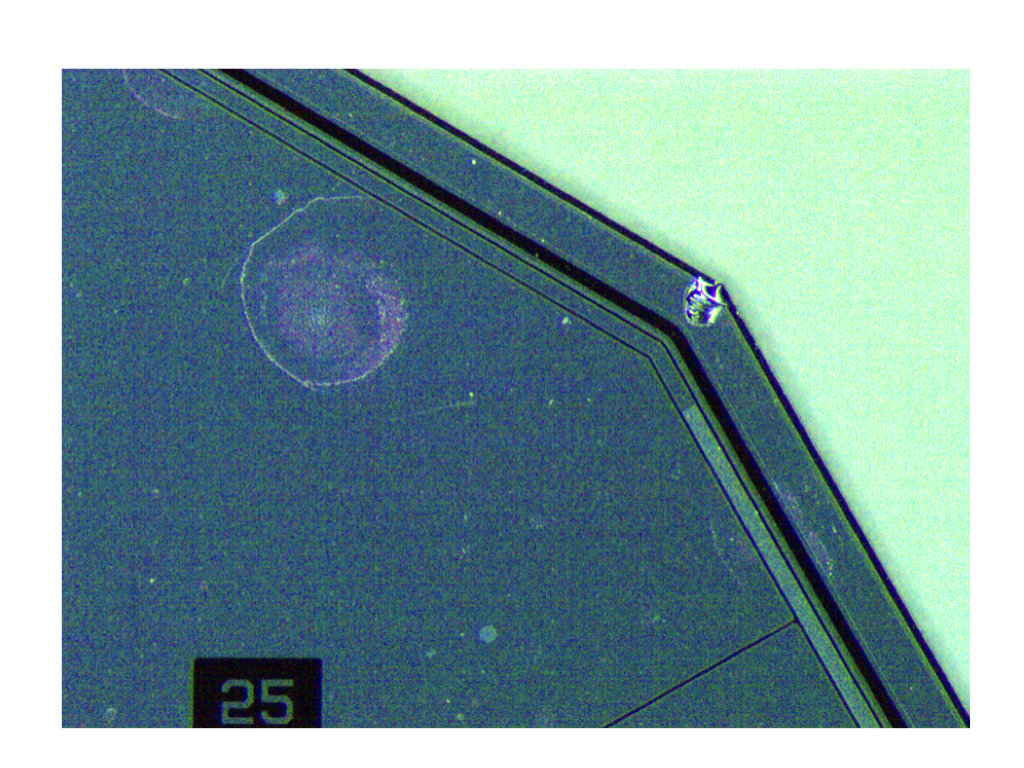}
		\subcaption{}		
	\end{subfigure}	
	\caption{Three example images of a sensor's surface, one with no defects (a),
	one with superficial scratches and stains (b),
	and one with a chipped edge (c).
	}
	\label{fig:scratch_examples}
\end{figure}

Here,
we present a case study in which deep learning techniques have been implemented at various construction
steps of silicon detector modules for the high granularity calorimeter
(HGCAL)~\cite{CERN-LHCC-2017-023} of the Compact Muon Solenoid experiment~\cite{cms:2008} at the Large Hadron Collider~\cite{evans:2008}. An HGCAL module is shown in Figure~\ref{fig:module}. As with many applications of silicon detectors, wire bonds between the segmented sensor pads and a circuit board are the primary mode of collecting signals. Figure 1b shows how wire bonds are arranged for HGCAL modules.   Visual inspection of the wire bonds for each HGCAL module is pivotal.  Moreover, the silicon sensor could have mechanical defects such as scratches,
chipped edges, or dust particles.  These defects might cause malfunctions such as low breakdown voltages.  Examples of sensor defects are shown in Figure~\ref{fig:scratch_examples}. The HGCAL will consist of roughly 30,000 sensors and tens of millions of wire bonds, all of which are to be QC tested.  Hence, the need to automate QC tasks for the HGCAL is essential.  

This work focuses on two proof-of-concept studies in which assembled modules and bare silicon sensors for the HGCAL undergo automated inspection.  In Section~\ref{sec:ttu}, we will describe the development of convolutional neural networks (CNNs)~\cite{INDOLIA2018679,SHARMA2018377,ageron2019hands-on}
 for categorizing wire bonding areas before and after wires are bonded.  We will discuss the performance of these algorithms and the prospects for optimizing feature detection efficiency at the cost of higher false-negative (where a bad image is identified as a good image) rates.  Then, in Section 2.2, we will describe the use of autoencoders for anomaly detection.  The anomalies here are physical defects in bare silicon sensors.  Finally, we will briefly discuss long-term objectives for QC automation in Section~\ref{sec:cern}, followed by a summary in Section~\ref{sec:summary}.

%% file: ttu.tex
\subsection{Image Classification using CNN}
\label{sec:ttu}
\begin{figure}[!htb]
  \centering
    \includegraphics[width=.8\textwidth]{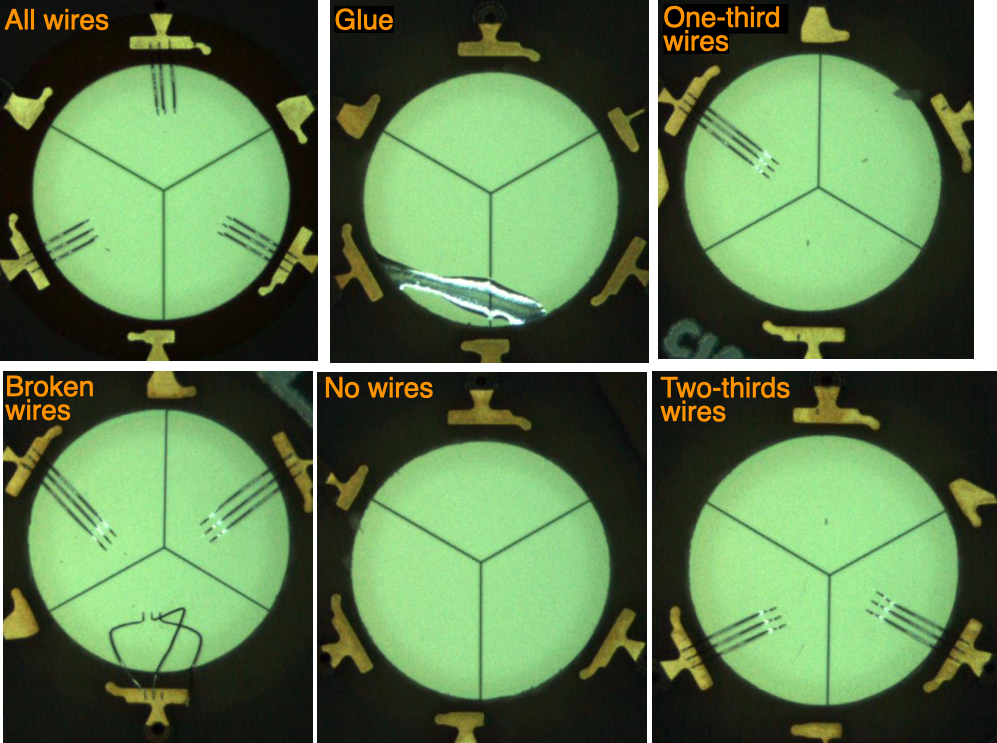}
  \caption{Images showing six classification categories. After the assembly steps but before wire bonding, no wires and no glue should be present.  After wire bonding, the three sections of the green pie should have three wire bonds each.  These three sections correspond to three distinct detector cells.  }
  \label{fig:label}
\end{figure}

For characterizing bonding areas, a CNN was trained to classify images.  A data set was collected consisting of 705 images.    Images were recorded before and after wire
bonding, and in some cases, after destructive tests of wire bond strengths.  Acquiring images at various stages of the assembly helped to produce training samples for missing wires and broken wires that could otherwise be rare occurrences.  Each image was labeled manually into six categories:  no wires, glue, broken wires, one-third, two-thirds, and all wires.  The absence of glue (no wires) and all wires category represent good-quality bonding areas before and after wire bonding, respectively.  The images labeled glue, broken wires, one-third, and two-thirds are considered poor-quality.  Epoxy can contaminate the bonding area during dispensing or when the circuit board is placed on the sensor.  Broken wires could be due to bad bonds.  Example images from each of the six categories are shown in Figure~\ref{fig:label}. Three unique subsets of images are formed, training (70\%), validation (20\%), and test (10\%).  Each subgroup is created with the images from each label category.  Data augmentations~\cite{shorten2019}, such as horizontal/vertical flips and rotations, are used to increase the data set to 4250 images.  


\begin{figure}[!htb]
     \centering
     
     \begin{subfigure}[b]{0.216\textwidth}
         \centering
         \includegraphics[width=\textwidth]{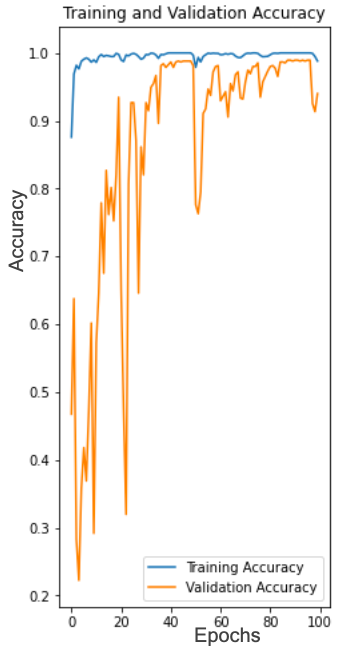}
         \caption{}
         \label{fig:acc}
     \end{subfigure}
     \hfill
    \begin{subfigure}[b]{0.21\textwidth}
         \centering
         \includegraphics[width=\textwidth]{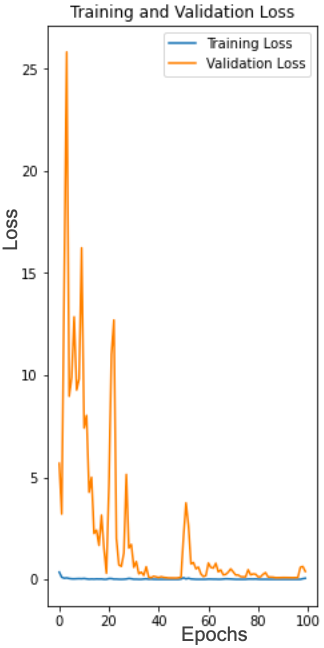}
         \caption{}
         \label{fig:loss}
     \end{subfigure}
     \hfill
     \begin{subfigure}[b]{0.5\textwidth}
         \centering
         \includegraphics[width=\textwidth]{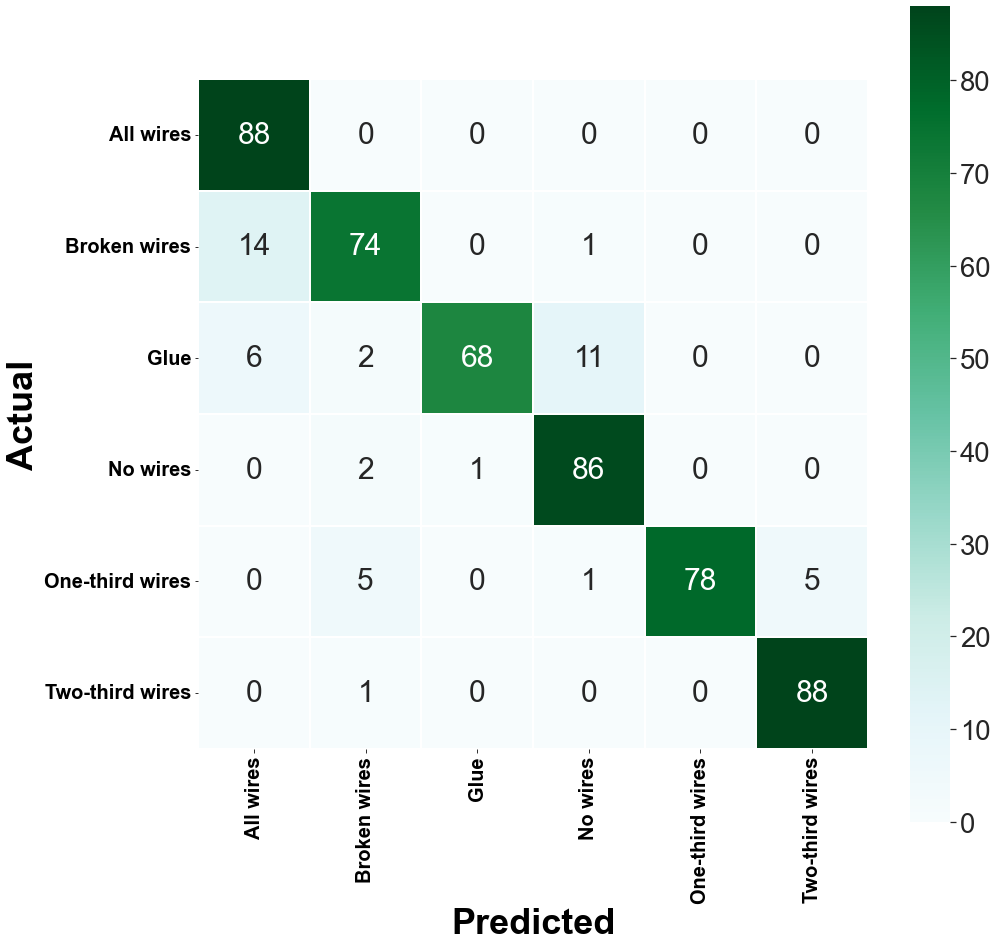}
         \caption{}
         \label{fig:cm-multiclass}
     \end{subfigure}
        \caption{The accuracy (a) and loss values (b) in the training and validation data sets. (c) 
    The confusion matrix representing the performance of the model on test data sets. The confusion matrix for the test data set has the actual (predicted) label on the vertical (horizontal) axis.}
        \label{fig:TL-performance}
\end{figure}

In scenarios where the training data set is not large enough, the transfer learning (TL) technique~\cite{yosinski2014transferable, transferlearning01} is often advantageous to use.  Moreover, the use of TL also helps reduce
the computational time compared to the time for training a model from scratch.  We used Keras~\cite{chollet2015keras} and Tensorflow~\cite{tensorflow2015-whitepaper} to implement the MobileNetV2~\cite{sandler2019mobilenetv2} model as a basis for TL.   Typically, we retrain the last 25 to 45 layers of the total 156 layers in the MobileNetV2 model.   Figures~\ref{fig:acc} and~\ref{fig:loss} show the accuracy and loss values, respectively, in the training and
validation data set for different epochs.  Figure~\ref{fig:cm-multiclass} shows the confusion matrix~\cite{pedregosa2018scikitlearn,kulkarni2021foundations,ageron2019hands-on} representing
the model's performance in the test data set.  Here, the loss is a  measure of the model's ability to make accurate predictions.  As seen in Figure~\ref{fig:acc}, the accuracy for the validation (training) data
set is as large as 98 (100)\%, and the loss shown in Figure~\ref{fig:loss} are close to 0 for both data sets.  We see a strong correlation
between the actual and predicted labels, demonstrating the model's potential.
The overall accuracy (true-prediction rate) which is the ratio of the sum of the diagonal entries to the total is about 91\%.  Furthermore, the precision, recall, and a $F-1$ score~\cite{grandini2020metrics} are evaluated for each label and are summarized in Table~\ref{tab:multiclass-report}.  A precision metric is computed from the ratio of the true positive (TP) occurrences to the total number of the positively predicted occurrences (TP and false positives (FP)).  The precision assesses the accuracy of the positive predictions.  A recall metric is computed from the ratio of the TP occurrences to the total number of positively classified units (TP and false negative (FN)).  The recall metric quantifies the fraction of positives correctly identified.  The $F-1$ score is a harmonic mean of the precision ($P$) and recall ($R$):  $F-1=\frac{2RP}{R+P}$.  An $F-1$ score of 1 (0) is the best (worst) score. For a given label,
the horizontal and vertical off-diagonal entries in Figure~\ref{fig:cm-multiclass} represent the FN and FP occurrences, respectively.

The results shown in Figure~\ref{fig:TL-performance} and Table~\ref{tab:multiclass-report} represent multi-class classification (one versus the rest).  However, some samples have more than one feature present in them.  For example, some images have glue present and broken wires.  To account for the fact that images can represent multiple classes, we investigated the use of multi-label classification.  

\begin{table}[h!]
\centering
\caption{Summary of the classification report for the multi-class classification.}
\begin{tabular}{llll}
\hline
Labels           & precision     & recall &  $F-1$ score     \\ \hline
All wires        &  0.81481 & 1.00000 & 0.89796 \\
Broken wires     &  0.88095 & 0.83146 & 0.85549 \\
Glue             &  0.98551 & 0.78161 & 0.87179 \\
No wires         &  0.86869 & 0.96629 & 0.91489 \\
One-third wires  &  1.00000 & 0.87640 & 0.93413 \\
Two-thirds wire  &  0.94624 & 0.98876 & 0.96703 \\
\hline           
\end{tabular}

\label{tab:multiclass-report}
\end{table}

The same image data set from the multi-class classification was relabeled in preparation for multi-label classification.  New training, validation, and test data sets were created from the relabeled images.  While relabeling, each instance is assigned to one or more classes, depending on the class attributes it possesses.  The same MobileNetV2 model used in multi-class classification is trained with the new training data set.  Similar to the multi-class model, the multi-label model performance on the training data set is similar to its performance on the validation data set.  The performance on the test sample is summarized in Table~\ref{tab:multilabel-report}.  
In this case, we see that the precision of the model's performance is remarkably high, with fewer FN than in the multi-class case (see Table~\ref{tab:multiclass-report}).  These performance metrics are expected to improve further as our training data set grows.


\begin{table}[h!]
\centering
\caption{Summary of the classification report for the multi-label classification.}
\begin{tabular}{llll}
\hline
Labels           & precision     & recall &  $F-1$ score     \\ \hline
All wires        & 0.99029       & 0.96226   &  0.97608          \\
Broken wires     & 0.94444       & 0.98077   &  0.96226          \\
Glue             & 0.99194       & 0.98400   &  0.98795          \\
No wires         & 1.00000       & 0.86996   &  0.93046          \\
One-third wires  & 1.00000       & 1.00000   &  1.00000          \\
Two-thirds wires  & 0.98947       & 1.00000   &  0.99471          \\

\hline           
\end{tabular}
\label{tab:multilabel-report}
\end{table}

\begin{figure}[!htb]
  \centering
    \includegraphics[width=.417\textwidth]{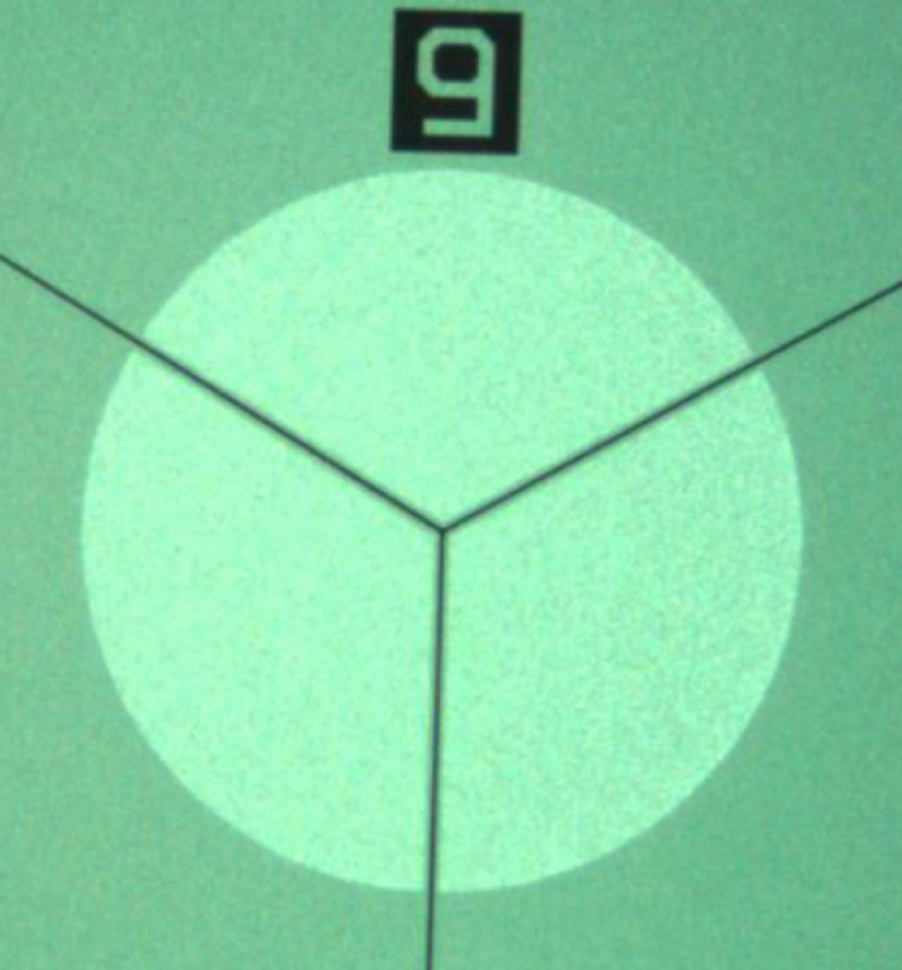}
    \includegraphics[width=.47\textwidth]{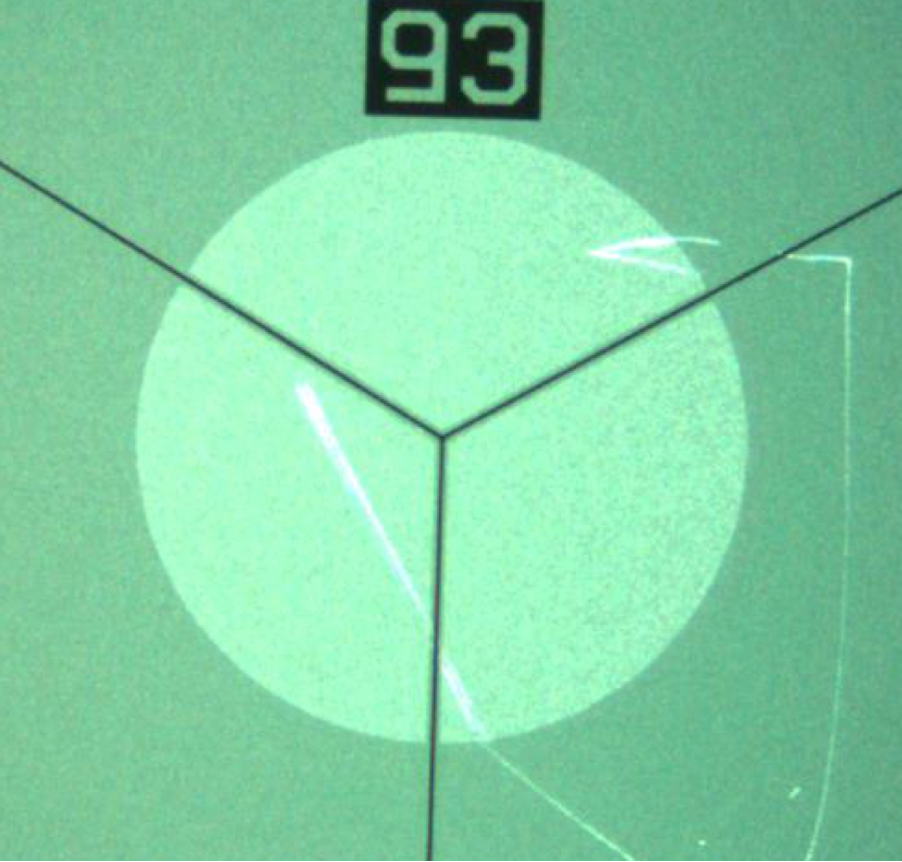}
  \caption{Images representing the no scratch (left) and scratch (right) category for classification of sensor qualities.}
  \label{fig:sensor_scratch}
\end{figure}

\begin{figure}[!htb]
  \centering
     \includegraphics[width=.7\textwidth]{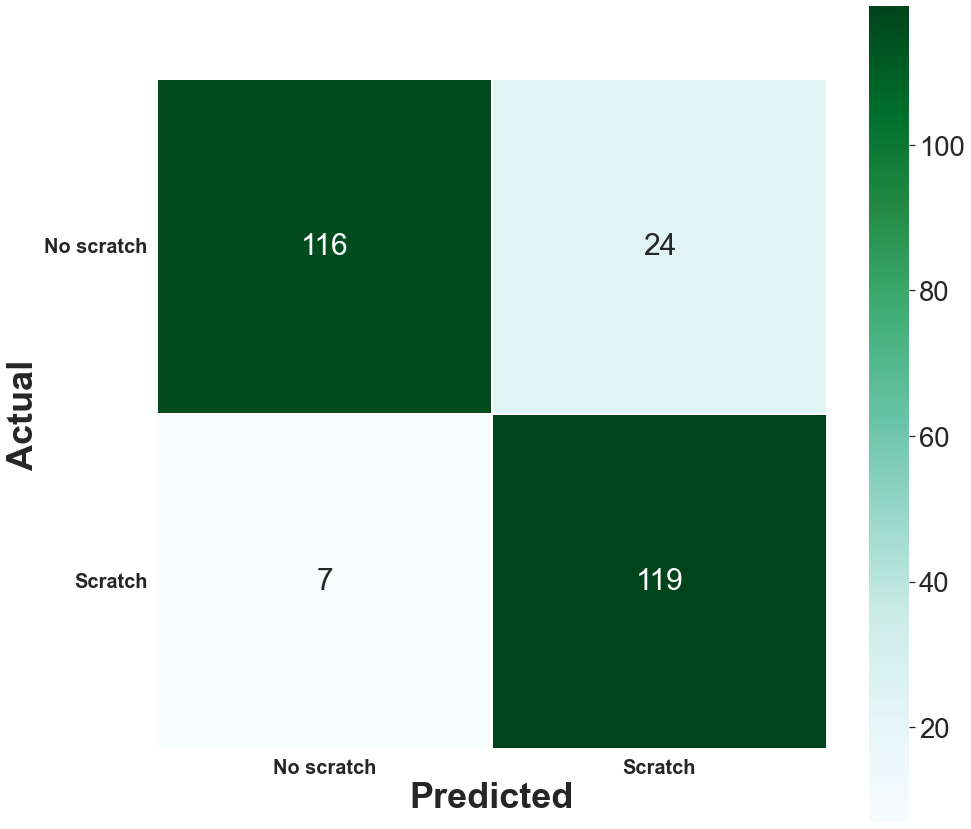}
  \caption{Confusion matrix showing the performance of the model on test data set. The vertical (horizontal) axis has the actual (predictied) entries for a given label.}
  \label{fig:cm-sensor}
\end{figure}

Similar supervised learning techniques were used to assess the quality of the silicon sensors by identifying physical defects such as scratches.  Sensor images having scratches or dust on the surface are labeled as `scratch' (positive class), and those without such flaws are labeled as `no scratch' (negative class).  An example image from each category is shown in Figure~\ref{fig:sensor_scratch}.  Since there are only two categories, a binary cross-entropy loss function is used.  The same transfer learning model, i.e., MobileNetV2, was used.  The model performance is presented in Figure~\ref{fig:cm-sensor} in which the TP, FP, and FN rates correspond to 88\%, 9\%, and 3\%, respectively.  


%% file: cern.tex
\subsection{Anomaly Detection using Autoencoder}
\label{sec:cern}

The use of AE in our study is inspired by many recent studies~\cite{PhysRevLett.121.241803,10.21468/SciPostPhys.6.3.030,PhysRevD.101.075021,PhysRevD.101.075042,PhysRevD.101.095004,knapp2020adversarially} on potential applications of novel deep-learning based methods for the detection of anomalous signals in LHC. The purpose of an AE is to mimic the identity function on a given set of input data.
However, the identity is approximate because of the AE's "bottleneck." 
The bottleneck is realized by first reducing the input data to lower dimensionality before reconstructing them back to their original dimensionality.
In practice, AEs are often designed as deep neural networks whose structure is adapted to the nature of the specific input data.
Due to their intrinsic translational invariance in the ability to detect features and to translate those into abstract but meaningful representations, CNNs~\cite{cnnlecun} are especially suitable for image-like data such as photos of silicon sensor surfaces.
Subsequently, the many thousands to millions of internal network parameters are determined by minimizing some reconstruction error metric evaluated on reference data in a process called "training."
AEs are often trained on the default data, i.e., without anomalies, when used for anomaly detection.
In the optimal scenario, the bottleneck's dimension is kept as small as possible while maintaining a sufficient reconstruction error.
Because the AE learns the identity function for anomaly-free data,  higher reconstruction errors will be observed with anomalous data. The error metric can thus be used as a tangible, discriminative observable indicating the presence of potential anomalies.

\begin{figure}
	\centering
	\includegraphics[width=1.\textwidth]{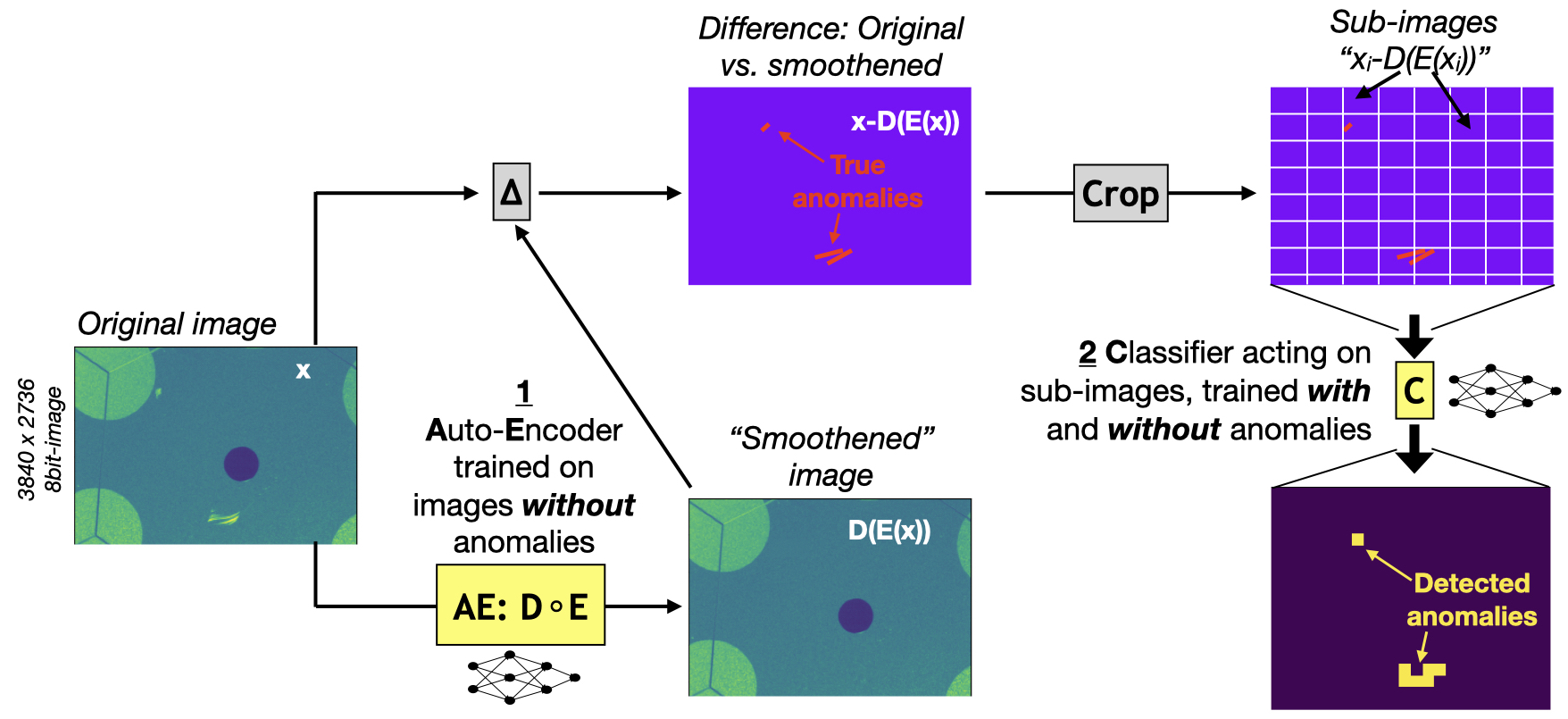}
	\caption{Strategy for the detection of visual anomalies in images of the surface of HGCAL silicon sensors inspected under the microscope.
	}
	\label{fig:strategy}
\end{figure}

\begin{figure}
	\centering
	\begin{subfigure}{0.49\textwidth}
		\includegraphics[width=1.\textwidth]{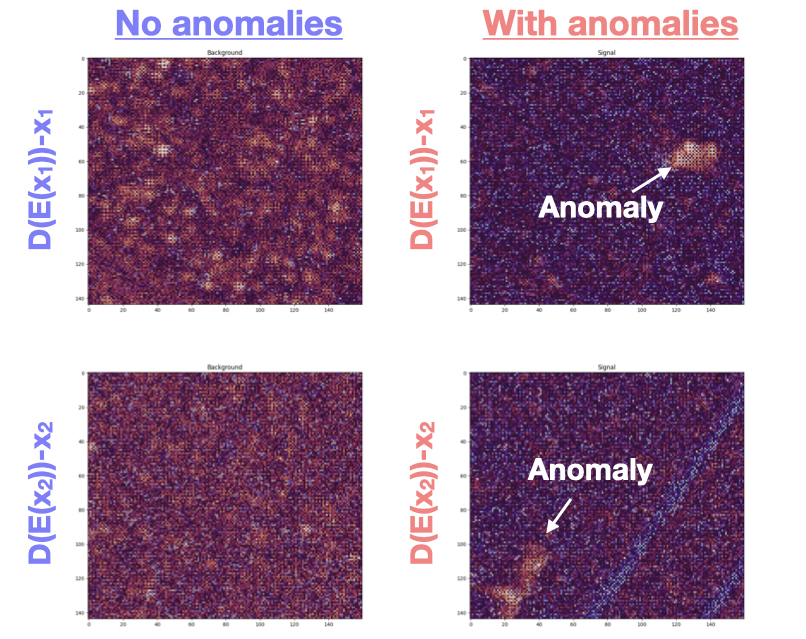}
		\subcaption{}		
	\end{subfigure}
	\hfill
	\begin{subfigure}{0.49\textwidth}
		\includegraphics[width=1.\textwidth]{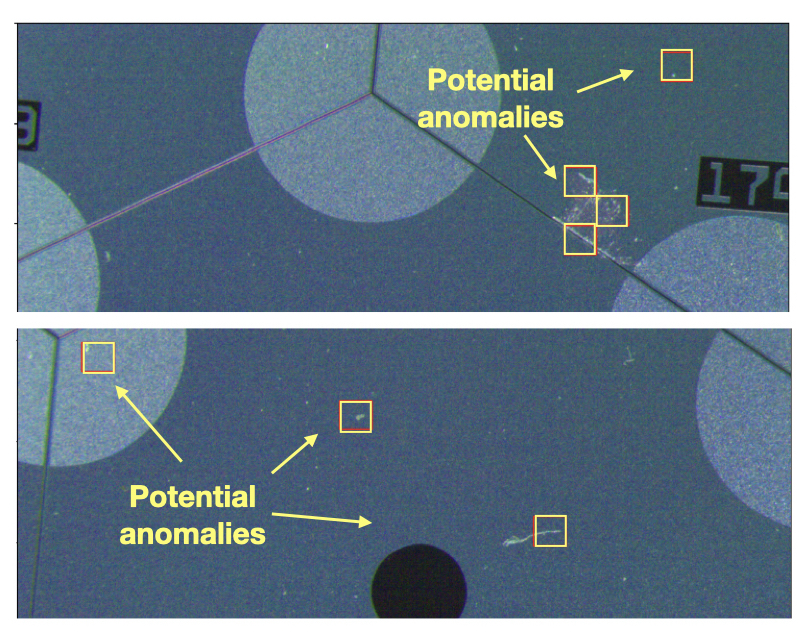}
		\subcaption{}		
	\end{subfigure}
	\caption{(a) Representative illustrations of reconstruction error maps from training images without and with visual anomalies. 
	The colour scale is arbitrary.
	(b) Application to new data. 
	The prototype algorithm presented in this section is able to detect regions of potential anomalies, but still misses a few.
	}
	\label{fig:proof_of_concept}
\end{figure}

An automated visual scanning setup has been developed as part of the electrical characterization of HGCAL silicon sensor prototypes at CERN. It was used to acquire $\mathcal{O}(10^4)$ high-resolution images of silicon sensor surfaces under magnification. Since those prototype sensors have been tested and handled extensively, a large variety of different anomalies (see Fig.~\ref{fig:scratch_examples}) on their surface could eventually be accumulated. 
A schematic of our approach is depicted in Fig.~\ref{fig:strategy} and consists of an ensemble of three CNNs. The first stage consists of an encoder (E) and a decoder (D) network that are evaluated in sequence and form an AE structure. This AE is trained exclusively on default images without anomalies. The difference between the original input image and the AE-processed images is computed in the second stage. Afterward, the full resolution image is cropped into smaller sub-images that are finally input sequentially into a classifier network.
It is the classifier's objective to reduce the two-dimensional reconstruction error map (see examples in Fig.~\ref{fig:proof_of_concept}) into a single discriminatory feature for the detection of anomalies. 
This ensemble of networks, each with $\mathcal{O}(10^5)$ free parameters, has been trained successfully with the aforementioned data set of HGCAL prototype silicon sensor images. The application of those networks after training is illustrated in Fig.~\ref{fig:proof_of_concept} as a proof-of-concept.

\subsection{Long-term Objectives}

Given the destructive effects of defects on the operation of sensors and wire bonds, the FN
rates of any QC algorithm should be as low as possible.  On the other hand, we consider
higher FP rates to be more acceptable.   In general, when developing multi-label classification algorithms, the threshold output discriminator for each class can be optimized for lower FN rates.   Threshold optimization was studied for the model presented in Section~\ref{sec:ttu}.  We found that the total number of FN examples could be reduced to one at the cost of an increased FP rate of a few percent.  The FN rate can be optimized for anomaly detection by incorporating more samples with anomalies than images of non-anomalous sensors to train the model.  In general, optimizing QC algorithms will require large quantities of images.  

For the construction phase of a detector, we target an anomaly detection efficiency of more than 99\%. 
Such a model will automate our inspection procedure; a preselection based on our model's inferences will save time and eye strain.  The model will also decrease the effect of human bias in anomaly detection, as it will no longer depend on the inspector's subjectivity. 
It must be noted that potential bias enters into the model from the labeling of the training images, which is currently foreseen to be done by humans.  The challenge is to consistently annotate defects over a large data set to start with a well-prepared and reliable training data set.
However, the limitation on the size of the training data set is because the number of samples with the defects (in both sensors and wire bonds) is currently small. 
In addition, in the case of sensor QC, where the data augmentation is yet to be applied, we argue that higher detection efficiencies can be achieved by augmenting the data set containing the defects before training the second classification model. 
Data augmentation will have to be designed such that it generates more defective images with annotated anomalies from the existing ones to extend the training data~\cite{shorten2019}. 
For instance, the data augmentation steps that would fit our circumstance are randomly cropping, rotating, and varying the brightness of the images.
These operations correspond to slight sensor misalignments (translational and rotational) and differing light exposures underneath the microscope. 
As a beneficial side-effect, the ultimate detection algorithm should be affected less by lightning conditions or camera position.  

We aim for the implementation of a model which can be updated and improved continuously when new data is available.  Incremental learning can be implemented if it can be assumed that the statistical properties of the defects will not change in the latest data (i.e., no concept drift)~\cite{widmer1996}.
Another approach for updating a model is a combination of incremental and transfer learning, where the old model is used as the starting point and its structure refined using the new data. 
By using this approach, new classes could be implemented into the model. 
Updating the model in this manner would be necessary due to new defects emerging during the evolution of the multi-year production phase of the silicon sensors or if other detector components were to be inspected in the future, for example.  No matter the ultimate choice, the ideal method for model updating would be one where minimal human effort is required.

Deep learning-based QC could also be a real-time computer vision application. 
In this form, the detection of flaws would be fast enough to be executed on a continuous stream of images.  This video-like processing would be similar to the face recognition feature on contemporary smartphones~\cite{chen2018}, and it would eliminate the need for the offline processing of stored images.

%% file: summary.tex
\section{Summary}
\label{sec:summary}

As the complexity of detector construction increases, the need for improved automation and capability in quality control (QC)  also increases.  Here we explore advanced computer vision algorithms built on deep-learning techniques to develop more automated and more capable tools for inspecting components and complete detector modules.   Several approaches were studied for analyzing images of silicon detector prototypes for the High Granularity Calorimeter being built to replace the Compact Muon Solenoid's endcap calorimeters.  This detector will have millions of channels and tens of millions of wire bonds, necessitating rigorous QC to ensure the desired performance of each detector channel and the system's longevity.  All images were collected using automated, programmable gantries.  We present tools to inspect bare sensors and assembled modules before and after wire bonding sensors to electronics.  Convolutional neural networks (CNN) are used to classify bonding areas according to six quality classes.  The precision among the classes is found to be in the range of 94--100\%. We also developed CNNs to classify images of sensors as good- and poor-quality based on the presence of physical defects.  In this case the accuracy is 88\% while 3\% of images contain bad features but are identified as good images.  Autoencoders were designed to identify physical defects such as scratches and dust on sensors using anomaly detection techniques.  The computer vision algorithms we studied can be used for fully automated QC with minimal expert intervention.  

As our data sets grow, we anticipate the performance of our algorithms will further improve.  We will explore automated updates to network weights to ensure our QC processes continuously improve.  We also hope to develop these tools into real-time image processing tools.  Our proof-of-concept studies demonstrate the power of machine learning-based tools for QC testing and likely can be widely applied for silicon detector QC in many other experiments.